\documentclass[aps,prb,floats,twocolumn]{revtex4-2}

\usepackage{
  amsmath,
  amssymb,
  array,
  booktabs,
  braket,
  color,
  float,
  graphicx,
  scrextend,
  hyperref,
  ulem,
  comment
}

\usepackage[dvipsnames]{xcolor}
\usepackage[caption=false]{subfig}
\date{\today}
\begin{document}

\title{Stiffness and atomic-scale friction in superlubricant MoS$_2$ bilayers}
\author{Rui Dong}\email[Contact email address: ]{dongru@tcd.ie}
\affiliation{School of Physics, AMBER and CRANN Institute, Trinity College, Dublin 2, Ireland} 
\author{Alessandro Lunghi}
\affiliation{School of Physics, AMBER and CRANN Institute, Trinity College, Dublin 2, Ireland}
\author{Stefano Sanvito} 
\affiliation{School of Physics, AMBER and CRANN Institute, Trinity College, Dublin 2, Ireland} 

\begin{abstract}
By using {\it ab-initio}-accurate force fields and molecular dynamics simulations we demonstrate that the layer stiffness
has profound effects on the superlubricant state of two-dimensional van der Waals heterostructures. These are engineered
to have identical inter-layer sliding energy surfaces, but layers of different rigidity, so that the effects of the stiffness
on the microscopic friction in the superlubricant state can be isolated. A twofold increase in the intra-layer stiffness
reduces the friction by approximately a factor six. Most importantly, we find two sliding regimes as a function of the 
sliding velocity. At low velocity the heat generated by the motion is efficiently exchanged between the layers and
the friction is independent on whether the sliding layer is softer or harder than the substrate. In contrast, at high velocity
the friction heat flux cannot be exchanged fast enough, and the build up of significant temperature gradients between the 
layers is observed. In this situation the temperature profile depends on whether the slider is softer than the substrate.
\end{abstract}

\maketitle

% INTRODUCTION
When the lateral forces between two sliding surfaces vanish or become extremely small, the surfaces 
are said to be superlubricant~\cite{Hirano1993}, a state often defined by a friction coefficient smaller than 
0.01. Structural superlubricity~\cite{M_ser_2004} is a particular superlubricant situation that emerges 
between dry and flat surfaces with incommensurate lattices. This structural peculiarity drastically suppresses 
the corrugation of the inter-layer sliding energy surface (ISES), so that the relative motion can take place 
with very limited energy dissipation. The ideal conditions for structural superlubricity are found when: i) the 
two surfaces are extremely rigid, so that elastic deformation is prevented over long length-scales; and ii) the 
surface-to-surface interaction is weak~\cite{VanDenEnde2012}. 

Two-dimensional (2D) materials, such as graphene, h-BN and transition metal dichalcogenides, are characterised 
by strong in-plane covalent bonds and weak inter-layer van der Waals interaction. Thus, vertically stacked 
2D heterostructures appear as an ideal materials platform for structural superlubricity. 2D compounds have 
been used as solid-state lubricants since many years~\cite{Martin1993,Rapoport1997}, although only recent advances 
in atomic-force microscopy have given us a thorough microscopic understanding of the superlubricity
phenomenon~\cite{Dienwiebel2004}. One can now find experimental demonstrations of structural superlubricity 
in graphene~\cite{Liu2012,Yang2013,Koren2015,Qu2020}, MoS$_2$~\cite{Li2017,Acikgoz2020,Vazirisereshk2020},
and in heterogeneous structures, h-BN/graphene~\cite{Song2018} and WS$_2$/graphene~\cite{Buch2018,Tian2020}.

Among the possible theoretical strategies to study microscopic tribology~\cite{Vanossi2013} the ``quasi-static'' 
approach has enjoyed significant popularity. This consists in computing the frictional forces from the gradient of the 
ISES~\cite{Koren2016}, which in turn can be obtained with {\it ab-initio} methods.~\cite{Hod2013} A variation of the same 
approach monitors the energy and forces during the movement of a ``slider'' over a ``substrate'', when the layers' internal 
degrees of freedom are either kept frozen~\cite{DeWijn2011,Leven2013} or allowed to relax~\cite{Reguzzoni2012}.
This is a good solution when looking at effects involving extended defects, such as grain boundaries, which require 
large simulation cells~\cite{Gao2021}. In general, the quasi-static approach works well in systems where the ISES is 
deep and the friction is dominated by the slider center of mass (CM) scattering. However, in a superlubricant situation the 
external forces dissipate into the internal atomic motion, a process that requires a molecular dynamics (MD) approach.
In this case the most typical setup consists in attaching the slider to a moving support through harmonic springs 
and in monitoring the spring forces over the MD trajectories~\cite{Ouyang2016,Wei2020,Dong2020}.
Empirical force fields are usually employed in this approach.

Here we use highly accurate machine-learning force fields, together with MD simulations without external driving 
forces, to answer a simple but crucial question: does the stiffness of the sliding layer affect the friction of a 
superlubricant system? We find that indeed this is the case, a twofold increase in the intra-layer stiffness
reduces the friction by approximately a factor six. Most importantly, the stiffness mismatch between the slider
and the substrate determines the thermal coupling between layers and the heat dispersion dynamics, resulting in 
different friction regimes at different sliding velocities.

Our simulations are for MoS$_2$ bilayers. The in-plane forces are described by a spectral neighbor analysis potential 
(SNAP)~\cite{SNAP}, generated with a procedure described before~\cite{AleSNAP}. This delivers a total-energy accuracy 
of 1.8~meV/atom, namely the MoS$_2$ potential energy surface is almost identical to the density-functional-theory (DFT) 
one, which is used to fit the SNAP. A simple Lennard-Jones potential extracted from van der Waals DFT calculations~\cite{TS2009} 
is employed for the inter-layer interaction. This returns a bilayer binding energy of 32.1~eV/\AA$^2$ for the 2H order, which 
is close to that computed by DFT, 35.7~eV/\AA$^2$. 
Similar results are obtained for other bilayer polymorphs. In addition, we generate two more SNAPs, obtained by rescaling the 
total energy of the distorted configurations included in the training set by either a factor 2 or a factor 1/2 (the total energy is 
measured from that of the equilibrium configuration). The resulting potentials have all the same energy minimum, namely the 
MoS$_2$ equilibrium geometry, but simulates materials with different stiffness (re-scaling the forces by a factor $\alpha$ 
changes the elastic tensor by the same amount). Layers described by these three SNAPs are denoted as {\it normal} 
(N), {\it soft} (S) and {\it hard} (H), respectively. Importantly, note that any bilayer constructed from these SNAPs has the same 
ISES. This not only allows us to perform dynamic simulations of friction in real materials with DFT accuracy, but also to isolate 
the effect of the in-plane stiffness on the superlubricity from those associated to the corrugation of the ISES.

\begin{figure}[htb] 
\begin{center}
\includegraphics[width=0.48\textwidth]{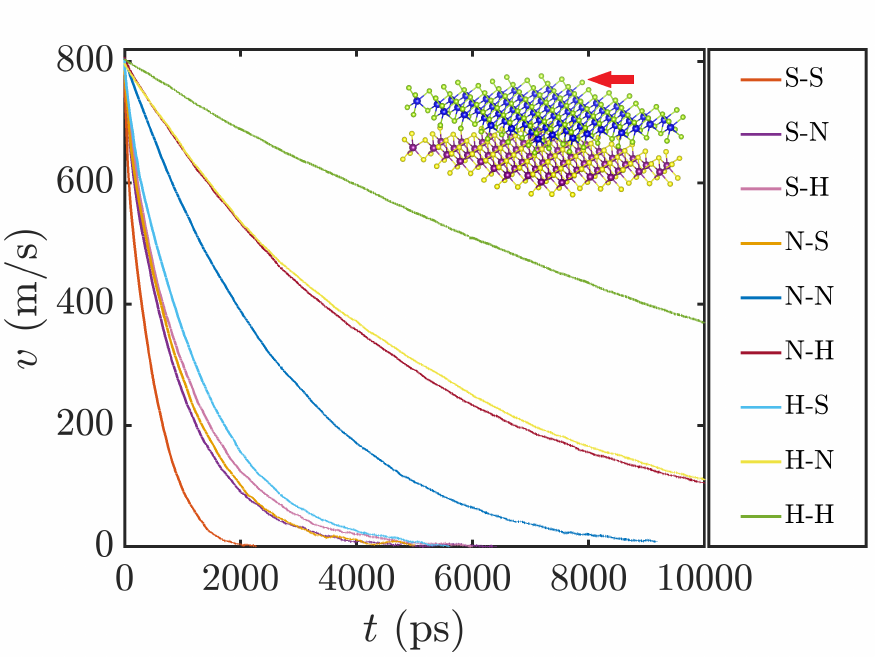}
\end{center}
\caption{(Color online) Time evolution of the velocity of the slider center of mass for the 9 bilayer types 
investigated. Inset: A ball-and-stick representation of the system used in the MD simulations, 
where a slider MoS$_2$ monolayer (top) moves above a MoS$_2$ substrate (bottom). Color code: 
Mo = blue/purple, S = gree/yellow.}
\label{FIG1}
\end{figure}

% METHODOLOGY
Our elemental structure is a $\sqrt{7}\times\sqrt{7}$ cell, in which the superlubricant state is obtained by
applying a twist angle of 21.8$^\circ$ between the two layers from the 2H configuration. This is the smallest 
strain-free cell that can be constructed~\cite{Dong2021}, returning a ISES corrugation of less than
0.1~meV/\AA$^2$. The MD simulations are then conducted over a $6\times6$ supercell of the $\sqrt{7}\times\sqrt{7}$
cell with periodic boundary conditions. One of the layer, the substrate, is kept fixed by removing the linear 
momentum of its CM at every MD step, while the other, the slider, is set in motion with an initial 
velocity of $v_0$=800~m/s (see Fig.~\ref{FIG1}). This rather high initial value (a piston in a combustion engine
moves at $\sim$25~m/s) gives us a velocity range large enough to converge well the friction-vs-velocity curve.
Note that our very flat ISES makes the slider performing Brownian motion at $\sim$30~m/s in equilibrium at room
temperature.
The system is equilibrated at 300 K using a Nos\'{e}-Hoover thermostat for 1~ns. After equilibration, snapshots 
are taken every 0.5~ns and used as the starting point for the sliding. With this set up the slider moves freely 
on the substrate, while its CM velocity is measured. 
Note that there is no constraint on any part of the slider, a fact that ensures the internal atomic vibrations not to be altered. 
During the sliding process the temperature of the substrate is thermostated at 300~K, but the temperature of slider is not 
controlled. This mimics experiments in vacuum, where the slider can only exchange energy with the substrate. Bilayer types 
are denoted as $\alpha$-$\beta$ ($\alpha, \beta=$~S, N and H) with $\alpha$ ($\beta$) defining the slider (substrate). For 
each configuration multiple MD runs are carried out to reduce the noise, and the analysis will be performed over the averaged
trajectories. 
%

% RESULT
Fig.~\ref{FIG1} shows the time evolution of the slider CM velocity for the nine possible bilayers. 
Clearly, the layer stiffness has a significant effect on the friction, since it takes only 2~ns to stop the soft MoS$_2$ 
bilayer (from its initial velocity of 800~m/s), while the normal and hard MoS$_2$ bilayers take ${\sim}$10~ns and 
${\sim}$40~ns, respectively. When combining monolayers of different stiffness, the general trend is maintained 
with the harder combinations preserving the motion for longer times. A minor anomaly appears for the combination 
with the largest stiffness mismatch between the layers, since the curves for H-S and S-H are not identical. We will 
come back on this point.

\begin{figure}[htb] 
\centering
\includegraphics[width=0.48\textwidth]{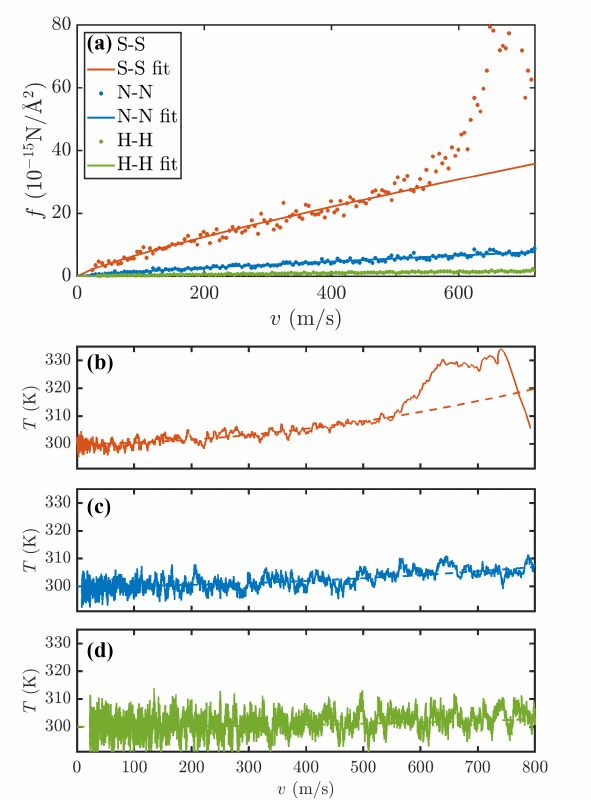}
\caption{(Color online) In the upper panel we present the friction force, $f$, as a function of the velocity for 
S-S, N-N and H-H MoS$_2$ homo-bilayers, where the circles are for the MD data and the lines are a fit to 
Eq.~(\ref{f_fit}). In the lower three panels we show the slider temperature as a function of the velocity, again 
for (b) S-S, (c) N-N and (d) H-H. In these, the solid lines are from MD, while the dashed ones are calculated 
by mean field according to Eq.~(\ref{gap}).}
\label{FIG2} 
\end{figure}

The acceleration-velocity curve, $a(v)$ can be obtained by simple finite-difference differentiation of the 
$v(t)$ curve over 5~m/s steps. Thus, the frictional force per unit area can be computed as $f(v)=\rho_s a(v)$, with 
$\rho_s$= 18.84 a.u./\AA$^2$ being the 2D density of the MoS$_2$ monolayer. Such quantity is presented in 
Fig.~\ref{FIG2}(a) against the slider velocity for the soft (S-S), normal (N-N) and hard (H-H) bilayers. 
We empirically find that the frictional force can be accurately fitted to a general expression
\begin{equation}
	f=-{\eta}(v/v_\mathrm{ref})^k\:,
	\label{f_fit}
\end{equation}
where $v_\mathrm{ref}$ is a reference velocity (set to 1~m/s), $\eta$ is the so-called viscous coefficient and 
$k$ is a parameter. Thus, in the range of velocity explored here the friction force remains in between the Stokes' 
($f\propto v$) and the Coulomb's ($f=$~constant) limit. This means, that even in the superlubricant regime the 
bilayers are away from equilibrium, resulting in $k\leq1$.~\cite{Muser2011} We notice that for the soft bilayer, 
$f$ grows dramatically for $v>$550~m/s, when also the slider internal temperature has a steep increase [see 
Fig.~\ref{FIG2}(b)], an effect not found for the other curves up to 800~m/s. We then decide to fit the $f(v)$ profile 
to Eq.~(\ref{f_fit}) only up to a maximum velocity, $v_\mathrm{max}\sim$~550~m/s, and the corresponding 
results are reported in Table~\ref{tab:1}.

Let us concentrate first on the homo-bilayers. In general, we find that as the rigidity of the system increases, 
$k$ approaches unity and the viscous coefficient reduces. In fact, $k=$~0.9 ($\eta=$~0.14) for the H-H bilayer, 
becoming 0.83 (1.02) and 0.82 (5.29), respectively for N-N and S-S. This behaviour is related to the ability of 
the slider to thermalise against the substrate, a feature that depends on the layer rigidity. In fact, from panels 
(b)-(d) of Fig.~\ref{FIG2} one can see that at any given velocity the slider temperature is larger for the softer 
layer, with H-H showing temperatures rather close to 300~K ($T$ of the substrate). The elevated temperature 
of the slider indicates that the CM kinetic energy is efficiently converted into internal thermal energy, leading to 
friction.
\begin{table}[!htb]
\centering
\caption{Parameters fitting the $f(v)$ curve to Eq.~(\ref{f_fit}). Here $\eta$ is the viscous coefficient in 
10$^{-6}$ m$\cdot$fs$^{-1}$s$^{-1}\cdot\rho_s$ ($\rho_s$= 18.84 a.u./\AA$^2$), $k$ is the velocity 
exponent and $v_\mathrm{max}$ is the maximum velocity considered in the fit. The fit correlation 
factor is $R$, while $G$ is the interfacial thermal conductance in units of MW/m$^2{\cdot}$K.}
\begin{ruledtabular}
\begin{tabular}{ c c c c c c c}
& $\eta$ & $k$ & $\eta k$ & $v_\mathrm{max}$ & $R$ & $G$ \\
\hline
{\bf S-S} & 5.29 & 0.82 & 4.34 & 500 & 0.95 & 83.3 \\
S-N & 1.17& 0.98 & 1.15 & 550 & 0.94 & 18.0 \\
S-H & 0.51 & 1.09 & 0.56 & 600 & 0.94 & 11.7 \\
N-S & 1.28 & 0.95 & 1.21 & 600 & 0.92 & 19.1 \\
{\bf N-N} & 1.02 & 0.84 & 0.86 & 750 & 0.95 & 49.1 \\
N-H & 0.26 & 0.96 & 0.86 & 750 & 0.92 & 5.88\\
H-S & 1.97 & 0.94 & 1.85 & 600 & 0.94 & 12.2\\
H-N & 0.20 & 1.00 & 0.20 & 750 & 0.93 & 5.80\\
{\bf H-H} &0.14 & 0.90 & 0.13 & 800 & 0.91 & 27.5\\
\end{tabular}
\end{ruledtabular}
\label{tab:1}
\end{table}

As mentioned before the soft bilayer displays an anomalous high temperature for $v\gtrsim550$~m/s, where the 
friction force abruptly deviates from Eq.~(\ref{f_fit}); namely, as the slider is put in motion at 800~m/s, its internal
temperature rapidly increases. This is because the friction is large and the vibration-energy flux injected into 
the slider exceeds the dissipative flux to the substrate. The latter is determined by the thermal coupling between 
the two layers. The elevated friction then causes a rapid reduction of the slider velocity, which in turn reduces the 
injected thermal flux. Thus, as the velocity drops, one reaches a steady-state situation in which the frictional 
energy injected is equal to the heat flow across the interface. Eq.~(\ref{f_fit}) is then restored. A similar behaviour 
is found also for the N-N bilayer at $v\geq$1000~m/s.

The low-velocity friction can be extrapolated from Eq.~(\ref{f_fit}) by taking $f\propto{\eta}kv$, where the product $\eta k$ is 
effectively the Stokes' friction coefficient. These are also reported in Table~\ref{tab:1} and clearly show a rather severe dependence 
of the low-velocity friction on the system rigidity. In fact, we find that doubling the stiffness leads to a friction reduction of approximately 
a factor six. Recalling that all the bilayers share an identical ISES, we conclude that the difference in friction is solely related to the 
layers internal dynamics and the thermal coupling between slider and substrate. Furthermore, by construction, all bilayers present 
an identical $\Gamma$-point breathing mode, which is associated to the rigid oscillation of the interlayer distance (see Fig.~S1 in 
the the supplemental material - SM). The kinetic energy associated to such mode, extracted from the vertical dynamics of the CM, does 
not change over the simulation time, meaning that it is not responsible for energy storing during the frictional motion. At the same time 
we observe little change in the interlayer distance, with the maximum increase of 0.015 \AA~found for S-S at $v>v_\mathrm{max}$. 
This, however, is consistent with the observed temperature profile and, therefore, it is not caused by scattering at the ISES.
\begin{figure*}[htb] %
\centering
\includegraphics[width=1.0\textwidth]{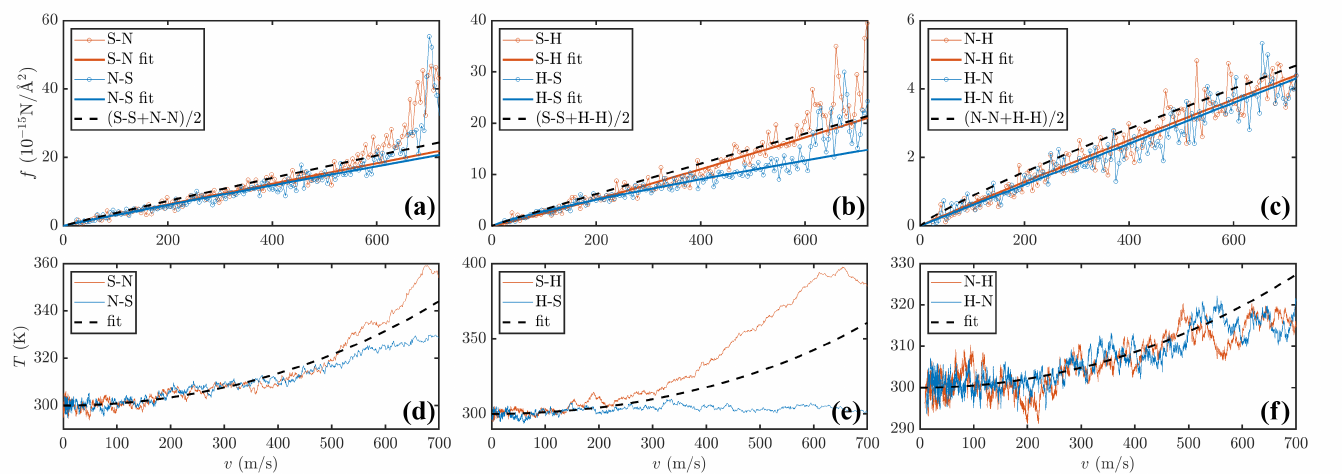}
\caption{Friction (upper panels) and slider temperature (lower panels) as
 a function of the slider velocity for the six different
hetero-bilayers investigated. The MD data (thin lines and circles) are fitted to Eq.~(\ref{f_fit}) (solid thick lines). The dashed lines
in the friction plots are obtained by averaging the friction fitted to Eq.~(\ref{f_fit}) for the corresponding homo-bilayers, while those 
in the temperature plots are calculated by using the prediction of Eq.~(\ref{gap}).}
\label{FIG3} 
\end{figure*}

Turning now our attention to heterostructures combining layers of different stiffness, Fig.~\ref{FIG3} shows both the $f(v)$ 
and $T(v)$ profiles for all the possible six combinations. Starting from S-N and N-S, it is clear that the friction is independent 
from the layer order for $v<$ 500~m/s, while above such velocity the soft slider is associated to a slightly larger friction.
A more noticeable difference can be found in the $T$ profile at high speed, where a soft slider moving on a normal substrate 
reaches 360~K, while a normal slider on a soft substrate does not exceed 330~K. More generally, the soft slider is constantly hotter 
than the normal one at any $v>$ 500~m/s. Notably, in our classical MD simulations at high $T$ monolayers of different stiffness
have the same heat capacity. This means that the different $T(v)$ profile for S-N and N-S must be associated to a different heat 
flux, which develops despite the rather similar frictional forces in the two cases. Such variation in heat flux becomes negligible 
as the velocity is reduced below $\sim$500~m/s. 

From Fig.~\ref{FIG3} we note that $f(v)$ computed for a hetero-bilayer is always smaller than the average friction of the associated 
homo-bilayers (dashed black lines), with the difference becoming more significant at high $v$'s. At the same time the temperature 
gap between the slider and the substrate (thermalised to 300~K) in hetero-bilayers is significantly larger than in homo-bilayers. 
These two facts together suggest that the thermal coupling between the layers in hetero-bilayers is weaker than in homo-bilayers. 
We can then model the friction as the sum of three contributions, $f=f_\alpha+f_\beta+f_{\alpha\beta}$, where $f_\alpha$ ($f_\beta$) 
is the friction originating from the energy dissipation to the slider (substrate) resulting from the ISES, while $f_{\alpha\beta}$ describes 
phonon-phonon scattering across the layers. Since the ISES is identical for all bilayers, $f_\alpha$ and $f_\beta$ must remain 
unchanged with the bilayer composition at a give $T$. Hence, $f$ of a hetero-bilayer remains lower than the average $f$ of the 
associated homo-bilayers, because $f_{\alpha\beta}$ is smaller for hetero-bilayers. This feature originates from the reduced overlap 
between the phonon spectra of monolayers of different stiffness (Fig.~S1 in SI).

An extreme situation is encountered for the S-H/H-S system [panels (b) and (e) of Fig.~\ref{FIG3}]. In this case the high-speed 
temperature increase is large for S-H (up to 400~K) and minimal for H-S, indicating that the main dissipation channel is through 
the soft layer. Since only the substrate is externally thermalised (as in vacuum experiments), such feature results in a two distinctly 
different $T(v)$ profiles, depending on the layers' order. By assuming $f_\mathrm{SH}$ and $f_\mathrm{H}$ to be much smaller 
than $f_\mathrm{S}$, and by comparing the low-$v$ $f(v)$ traces of the S-H and S-S systems, we can conclude that in 
the soft homo-bilayer $f_\mathrm{SS}$ contributes to about 30\% of the total friction, namely it is similar to $f_\mathrm{S}$ over the
entire $v$ range. Finally, the N-H and H-N hetero-bilayers show similar friction and temperature profiles, resembling the S-H case at 
low velocity. As the layers are both relatively rigid we do not note any significant heating at all velocities.

\begin{figure}[!htb] \
\centering
\includegraphics[width=.48\textwidth]{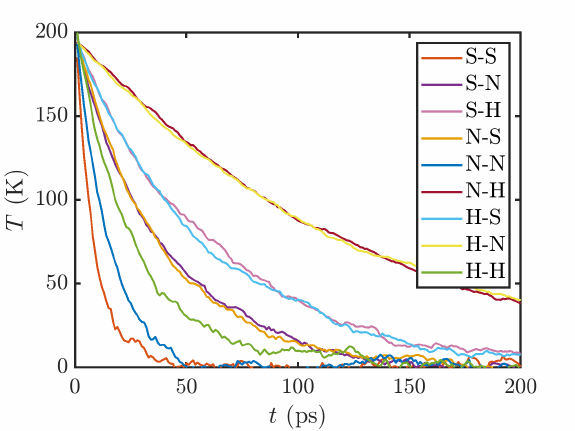}
\caption{(Color online) Time evolution of the temperature gap between the two layers of a bilayer, following equilibration
at 400~K and 200~K, respectively. Each curve is the average over 10 different trajectories.}
\label{FIG4} 
\end{figure}

Our analysis can be made more quantitative by determining the interfacial thermal coupling between the layers. Since the out-of-plane 
thermal conductivity of a bilayer is ill-defined, we instead compute the interfacial thermal conductance, $G$. This simply relates the 
heat flux, $q$, with the temperature difference ${\Delta}T$, $q=G{\Delta}T$, and it can be extracted from MD simulations. We first equilibrate the 
two layers at 400~K and 200~K, respectively. In the absence of external thermal reservoirs, the temperature difference between the two 
layers decays exponentially in time from its initial value, $\Delta T_0$=200~K (see Fig.~\ref{FIG4}), as
\begin{equation}
	\Delta T(t)= \Delta T_0\:\mathrm{exp}\left(-\frac{2G}{C_v}t\right)\:,
	\label{temperature relaxation}
\end{equation}
where $C_v$ is the specific heat. $G$ can then be extracted by monitoring the time-evolution of $\Delta T$. In performing the fit 
$C_v$=75.75 J/mol$\cdot$K$^{-1}$ is calculated from the total energy fluctuations. As expected $G$ is found not to depend 
on the direction of the heat flux and the computed values (per unit area) are reported in Table~\ref{tab:1}. 

The thermal coupling between layers depends on the relative strengths of the inter- and intra-layer interactions, namely the binding 
energy and the bond stiffness. Thus, heat is transferred across the layers when the motion of the atoms in one layer drives the motion 
in the other. As the inter-layer van der Waals forces are identical for all heterostructures, in this case only the layer stiffness differentiates
the thermal coupling. Thus, for homo-bilayers we find $G$ to be relatively large and decreasing as the layer stiffness increases.
In contrast, in hetero-bilayers, the interface conductance is determined by the ``acoustic mismatch'', namely by the relative thermal 
impedances, combined with the phonons coupling at the interface. These two factors together make the N-H combination being
the less conductive and the S-N the most. 

We are now in the position to complete our analysis. During the sliding process an heat flux, $vf$, is injected into the two layers. 
A steady-state situation is maintained when the heat transferred across the layers is constant and it is balanced by the heat flux 
to the thermostat in the substrate. In this situation we have
\begin{equation}
	{\Delta}T=\frac{vf(v,T)}{2G}\:,
	\label{gap}
\end{equation}
where $f\propto T^{1.6}$, a temperature dependence that has been obtained for the N-N bilayer by fitting Eq.~(\ref{f_fit})
at different temperatures [see Fig.~S2 in SM]. The calculated $T(v)$ profiles are then plotted in the panels (b)-(d) of Fig.~\ref{FIG2} and (d)-(f) 
of Fig.~\ref{FIG3} as dashed lines (when the substrate is thermostated then, $T(v)=\Delta T$). The curves show an excellent
agreement between the measured temperature and that provided by Eq.~(\ref{gap}), for both homo- and hetero-bilayers. This means
that, for $v<v_\mathrm{max}$, there is steady-state heat flux across the layers, which is then broken at higher velocities, for which 
the severe heating of the slider is observed.

%SUMMARY

In summary, by using {\it ab-initio}-accurate force fields we have investigated the superlubricant state of MoS$_2$ bi-layers. These 
have different in-plane stiffnesses, but identical inter-layer sliding energy surface, a feature that allows us to investigate the sole effect 
of the stiffness on the friction. In general, we find that the friction goes as $v^k$ with the exponent remaining close to unity
for rigid layers and deviating for soft layers and heterogeneous bilayers. For homo-bilayers a factor-two change in stiffness results in 
approximately a sixfold variation in friction. Hetero-bilayers follow a similar trend, although the low-velocity friction remains in general 
small. Similarly to other bilayer systems, we find that the out-of-plane motion of the layers is not a major energy dissipation channel.

The thermal coupling between the layers determines the heating during the sliding process. At low velocity a steady-state is establish,
where the temperature difference between the slider and the substrate sustains a constant heat flux. In this regime the friction and the
temperature of the slider are independent on whether the slider is softer or harder than the substrate. In contrast, at elevated temperature
the slider can heat up significantly, with the effect being much more pronounced for soft sliders. The crossover between these two regimes
depends on the specific layer combinations. In particular, we find that hetero-bilayers composed of rigid materials can sustain ultra-low
friction and moderate heat up even at extremely high velocities. 

This work has been supported by IMRA Europe S.A.S. and by Science Foundation Ireland (grant 12/RC/2278$_-$P2).
We thank St\'ephane Bourdais for discussion. Computational resources have been provided by the Trinity Centre 
for High Performance Computing (TCHPC).

\renewcommand{\baselinestretch}{1}

\end{document}